\documentclass[fleqn,10pt,twocolumn]{SICE_FES25}
\usepackage{amsmath}
\usepackage[font=small,labelfont=bf]{caption}
\usepackage[dvipsnames]{xcolor}
\usepackage{eso-pic}
\usepackage{graphicx}
\usepackage{url}
\usepackage{fancyhdr} 

\fancypagestyle{firstpagehead}{%
  \fancyhf{}%
  \fancyhead[C]{\footnotesize
    This work has been submitted to the IEEE for possible publication.
    Copyright may be transferred without notice, after which this version
    may no longer be accessible. This work has been accepted for the 2025 SICE Festival with Annual Conference (SICE FES 2025), September 9-12, Chiang Mai, Thailand.}%
}

\title{An AI-Based Shopping Assistant System to
       Support the Visually Impaired }

\author{Larissa R. de S. Shibata${}^{1\dagger}$, Ankit A. Ravankar${}^{1}$, Jose Victorio Salazar Luces${}^{1}$, and Yasuhisa Hirata${}^{1}$}
\speaker{Larissa R. de S. Shibata}

\affils{${}^{1}$Department of Robotics, Tohoku University, Sendai, Japan\\
(E-mail: {de.souza.shibata.larissa.rika.r4@dc.tohoku.ac.jp}, {\{ankit,j.salazar,hirata\}@srd.mech.tohoku.ac.jp})\\
}

\abstract{%
Shopping plays a significant role in shaping consumer identity and social integration. However, for individuals with visual impairments, navigating in supermarkets and identifying products can be an overwhelming and challenging experience. This paper presents an AI-based shopping assistant prototype designed to enhance the autonomy and inclusivity of visually impaired individuals in supermarket environments. The system integrates multiple technologies, including computer vision, speech recognition, text-to-speech synthesis, and indoor navigation, into a single, user-friendly platform. Using cameras for ArUco marker detection and real-time environmental scanning, the system helps users navigate the store, identify product locations, provide real-time auditory guidance, and gain context about their surroundings. The assistant interacts with the user through voice commands and multimodal feedback, promoting a more dynamic and engaging shopping experience. The system was evaluated through experiments, which demonstrated its ability to guide users effectively and improve their shopping experience. This paper contributes to the development of inclusive AI-driven assistive technologies aimed at enhancing accessibility and user independence for the shopping experience. }


\begin{document}

\maketitle
\thispagestyle{firstpagehead} 
\pagestyle{empty}             

\section{Introduction}
Shopping is a vital aspect of everyday life that extends beyond the acquisition of goods. It plays a role in personal identity, social participation, and the sense of autonomy. However, for individuals with visual impairments, shopping in physical environments introduces considerable barriers. Navigating stores, locating specific products, and making informed purchasing decisions are tasks heavily reliant on visual input, often requiring assistance from others. According to the World Health Organization (WHO), over 2.2 billion people globally are affected by visual impairments, highlighting the broad social importance of inclusive and accessible shopping solutions \cite{world2019world}.

In recent years, several assistive technologies have emerged to address these challenges\cite{ogura2023}. Mobile applications such as TapTapSee, Navilens, Seeing AI, and Vision utilize device cameras and voiceover functionalities to identify objects and texts\cite{serena2019innovative}. For navigation, GPS-based solutions including BlindSquare, VoiceMap, and SWAN offer route guidance, primarily in outdoor environments \cite{baker2006consumer}. Additionally, crowd-sourced platforms like ``Be My Eyes"\cite{bemyeyesHome} connect visually impaired individuals to volunteers or AI services for ad-hoc assistance. While these solutions have made significant contributions, they face notable limitations. Specifically, mobile apps often require manual control and familiarity with multiple interfaces, GPS-based solutions are unsuitable for precise indoor navigation, and remote human-based solutions may not offer real-time or contextually rich guidance.
\begin{figure}[!ht]
\begin{center}
\includegraphics[scale=0.17]{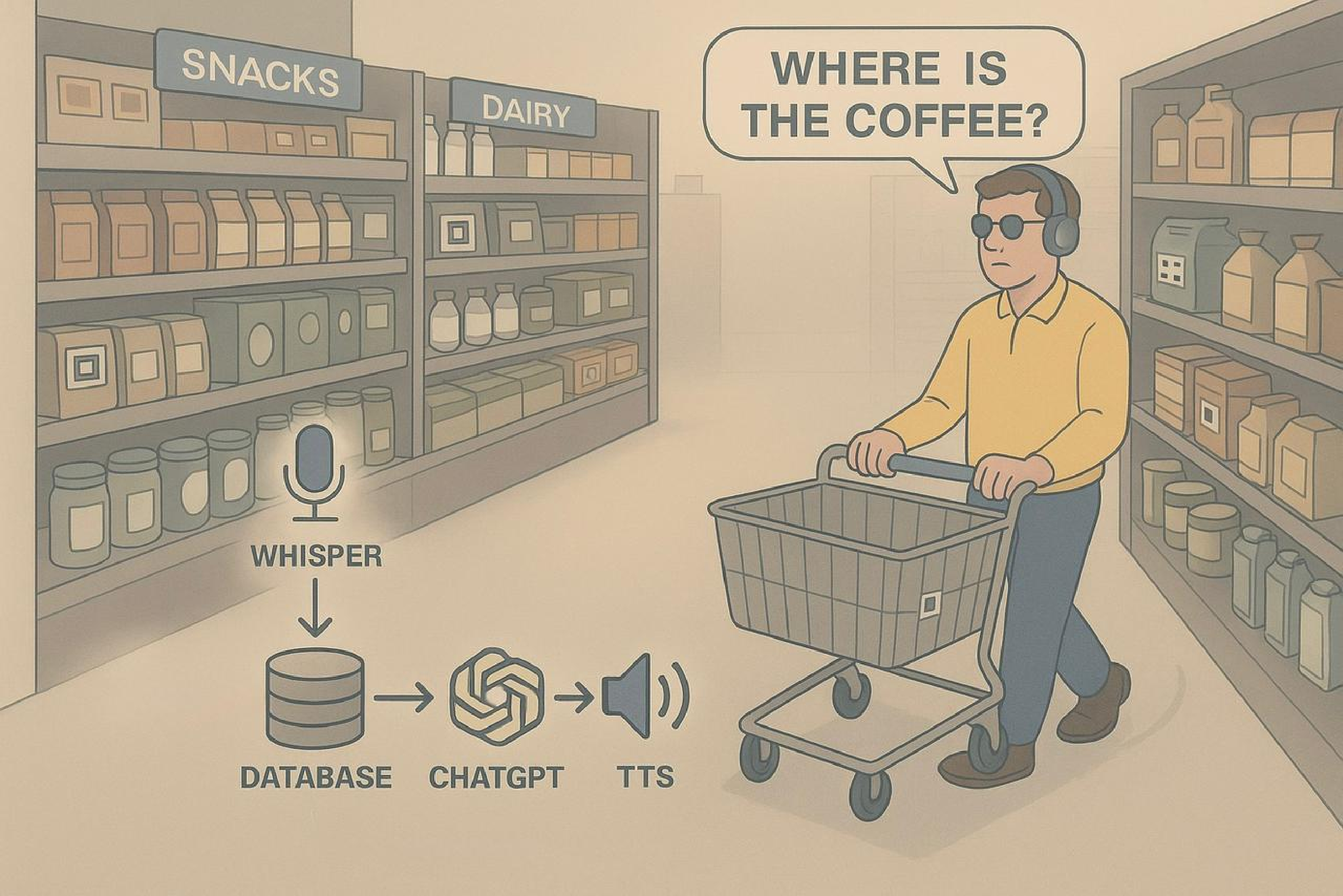}
\caption{\label{figure2} Concept image of our AI-based Shopping Assistant System for Visually Impaired. Our proposed system guides user to different sections in a store and provide real-time multi-modal information about the environment and products on the shelves.}
\end{center}
\end{figure}
More recently, robotics and AI-based systems have been explored to provide embodied assistance to elderly or visually impaired individuals \cite{ravankar2015connected,domingo2012overview,ravankar2022care,khan2020ai,ravankar2019itc}. Robotic guides and wearable navigation aids have been proposed to offer real-time environment feedback using sensors such as LiDAR, depth cameras, and computer vision techniques\cite{bamdad2024slam,lacey1998application,salazar2018path,liao2020human}. For example, autonomous robotic assistants have been trialed for guiding users in structured indoor spaces like museums and airports. Recent advancements in sensor technology, including multi-sensor fusion and autonomous robot mobility have brought significant progress in realizing smart and autonomous navigation for diverse applications \cite{alatise2020review,ravankar2020safe,thrun2002probabilistic,beomsoo2021mobile}. While promising, these systems are often specialized, infrastructure-dependent, or focused solely on navigation rather than comprehensive in-store support. Furthermore, indoor localization technologies, including RFID tags, BLE beacons, and vision-based approaches such as ArUco markers have been investigated for assistive applications, yet integrating them seamlessly with user interaction models remains an open challenge\cite{zafari2019survey}.

Furthermore, most existing systems approach shopping as a rigid and goal-oriented process, supporting only predefined queries or fixed routes. This perspective overlooks the dynamic and opportunistic nature of in-store shopping, where consumers often make spontaneous decisions influenced by promotions, store layouts, and cultural preferences\cite{kostyra2017food}. Such subtle and context-dependent behaviors remain largely inaccessible to visually impaired individuals using current technologies \cite{yuan2019constructing,tullio2021you,yu2015retail,balconi2024choose}.

To address these gaps, this work proposes a unified AI-based shopping assistant system embedded within a smart shopping cart. Unlike prior solutions that rely on disparate devices and fragmented functionalities, the proposed system integrates computer vision, automatic speech recognition (ASR), natural language interaction, and indoor localization using ArUco markers into a single platform. Through voice commands and multimodal feedback, users can explore store environments, identify product locations and details, and receive context-aware auditory guidance—all without switching devices or interfaces. This design aims to promote not only accessibility but also autonomy and spontaneity during shopping experiences. The main contributions of this work are summarized as follows:

\begin{itemize}
\item A unified shopping assistance platform that integrates environment description, product information retrieval, and navigation into a single embodied system tailored for visually impaired users.
\item An AI-driven interaction model combining Whisper ASR, ChatGPT for dialogue and contextual reasoning, and ArUco marker-based localization for precise indoor navigation.
\item User-centered experimental evaluation under normal and simulated visual impairment conditions, demonstrating system effectiveness and analyzing usability and workload through NASA-TLX metrics.
\end{itemize}

The remainder of this paper is organized as follows. Section 2 introduces the proposed system architecture and implementation details. Section 3 presents the experimental design, results, and analysis. Finally, Section 4 concludes with discussions on system performance, limitations, and future development directions.

\section{Proposed System}

This section introduces the AI-based assistive shopping system developed to enhance the autonomy of visually impaired individuals in supermarket environments. The system is designed to address common challenges such as locating store sections, retrieving product information, and navigating aisles. To provide these functionalities in an intuitive and accessible manner, the system integrates computer vision, speech recognition, natural language processing, and indoor localization into a unified smart shopping cart platform.
The proposed system supports three core user functions:

\begin{itemize}
\item \textbf{Environment Understanding and Awareness:} Delivering real-time auditory descriptions of store sections and surroundings.
\item \textbf{Product Information Retrieval:} Enabling users to query product details, including location and price, through voice commands.
\item \textbf{Indoor Navigation Assistance:} Guiding users to desired products or sections with step-by-step voice instructions.
\end{itemize}
To achieve these capabilities, the system architecture consists of five functional modules: \textit{Listening}, \textit{Capture}, \textit{Processing}, \textit{Navigation}, and \textit{Speaking}. The relationship among these modules is illustrated in Figure~\ref{figure2}.

\begin{figure*}[!ht]
\begin{center}
\includegraphics[width=\textwidth]{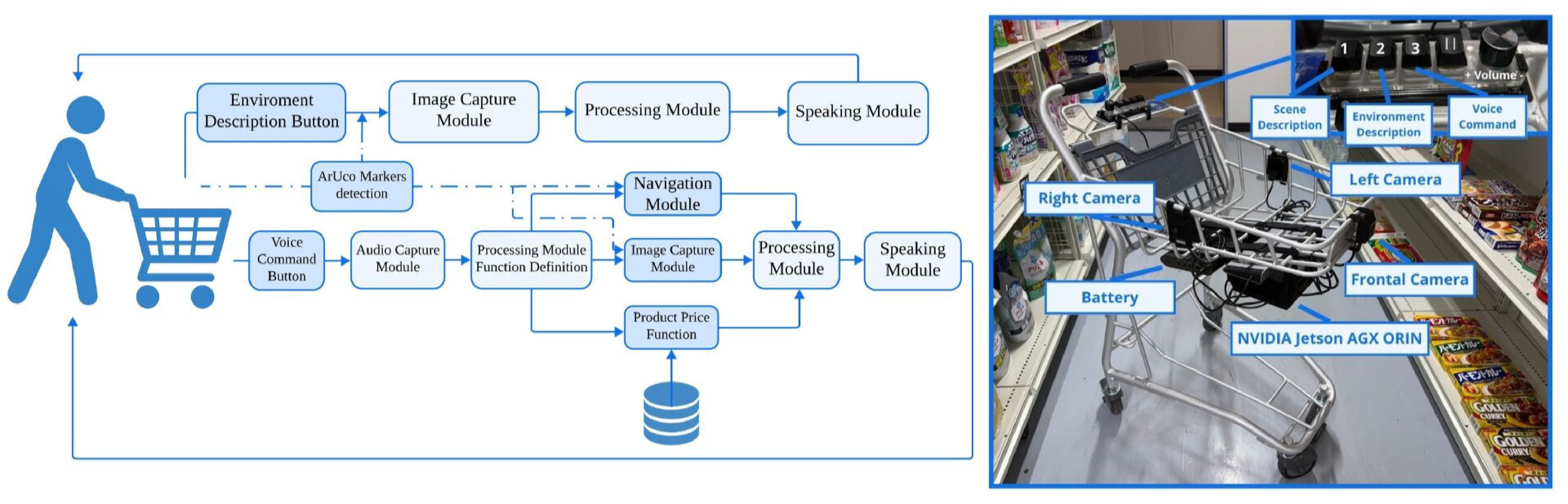}
\caption{\label{figure2} Architecture of the proposed system, including the image capture, audio capture, processing, navigation, and speaking modules. The developed smart cart with sensor integration is also presented. Inset image shows the button module developed for the user.}
\end{center}
\end{figure*}

\subsection{{System Operation Workflow}}

The system remains in an idle state within the \textit{Listening Module} until activated by the user via one of three physical buttons on the cart interface:
\begin{itemize}
\item \textbf{Environment Description Button:} Initiates the \textit{Image Capture Module}, which activates multiple onboard cameras to capture images from various angles and forwards them to the Processing module.
\item \textbf{Voice Command Button:} Triggers the \textit{Audio Capture Module}, allowing the user to dictate questions or requests. The audio input is transcribed into text using the Whisper speech recognition model. If necessary, this also activates the Image Capture and Processing modules sequentially.
\item \textbf{Section Description Button:} Provides real-time auditory feedback by utilizing ArUco marker recognition data. The recognized marker is interpreted to determine the user’s current section, and the information is announced via the Speaking module.
\end{itemize}

While in operation, the system’s cameras continuously monitor the environment for ArUco markers, independent of button presses. Detection of these markers plays a critical role in indoor localization, allowing the system to determine the cart’s precise position and orientation relative to store sections or shelves.

When user input is processed in the \textit{Processing Module}, the AI interprets the command and selects the appropriate action pathway. This may involve:

\begin{itemize}
\item Performing a semantic search in the local product database using text embeddings to retrieve relevant product information such as price or shelf location.
\item Sending images for analysis via a vision-language model integrated with ChatGPT to describe the environment or answer user-specific queries.
\item Accessing indoor navigation functions to calculate and initiate guidance to a target destination.
\end{itemize}

If navigation is required, the system activates the \textit{Navigation Module}. This module computes the current position using detected ArUco markers and plans a path to the target location. The path is translated into a sequence of human-understandable audio instructions, which are then relayed to the user through the \textit{Speaking Module}.

The Speaking Module utilizes a text-to-speech (TTS) engine to convert AI-generated textual responses or navigation instructions into natural-sounding speech. The audio output is transmitted through the user’s headphones to ensure clear communication without external disturbances. Upon completion of each response cycle, the system returns to the \textit{Listening Module} to await the next user interaction.

\subsection{{ArUco Marker Encoding and Database Structure}}

ArUco markers are fundamental to the system’s indoor positioning capability. Each marker encodes a unique identifier (ID) that corresponds to either a specific shelf or a designated location within the supermarket layout. These marker IDs are mapped to human-readable labels, such as ``Dairy,'' ``Snacks,’’ or ``Cleaning Products,’’ via a local reference table stored in the database.
The product database is structured to store comprehensive information, including:

\begin{itemize}
\item Product name, brand, and variety.
\item Shelf ID linked to the corresponding ArUco marker.
\item Vertical position on the shelf (from top to bottom).
\item Horizontal position on the shelf (from left to right).
\item Product price.
\end{itemize}

This structured organization enables accurate retrieval and delivery of product location and pricing information in response to user queries. By leveraging semantic search capabilities through text embeddings, the system can handle variations in user input and still return precise or close-matching product results.

\section{System Architecture}

{The proposed system adopts a modular and scalable architecture, where each functional block—Listening, Capture, Processing, Navigation, and Speaking—is implemented as an independent thread. These threads communicate and synchronize using shared queues and signaling mechanisms to ensure proper coordination and responsiveness. For example, when a user presses the voice command button, the Audio Capture thread captures and transcribes the audio, then passes the resulting text to the Processing thread through a shared communication buffer. Depending on the interpretation of the input, the Processing thread may trigger additional actions, such as image processing, product lookup, or navigation planning.}

{This multithreaded design enables parallelism while preserving modularity, making it possible to update or replace individual components without affecting the overall workflow. Each module operates autonomously and only activates when relevant input is received, ensuring efficient resource usage and low-latency responses.}

{Moreover, this architecture promotes the reusability of each module in other applications. For instance, the voice interaction and environment description threads can be integrated into service robots or public information kiosks for visually impaired users. Similarly, the navigation planning logic based on visual markers and route generation can be repurposed for indoor guidance systems in hospitals, libraries, or other indoor spaces. The system's thread-based modularity lays the groundwork for developing a more general methodology for integrating multi-modal AI components into a wide range of assistive and service-oriented technologies.}

\subsection{Image Capture Module}
The Image Capture Module is responsible for acquiring visual data from the supermarket environment and detecting ArUco markers for localization. This module plays a two important role: recognizing the user’s location and surroundings, and providing visual input for environment description when requested.

To achieve wide coverage and robust perception, the system employs three ELP-USBGS1200P01 cameras, each equipped with a 120-degree distortion-free lens. These cameras are strategically mounted on the left, front, and right sides of the shopping cart to capture a comprehensive view of the environment. We used the NVIDIA Jetson AGX Orin platform for collecting and processing the sensor data, doing all the computations, and running the AI modules. The system was running on Ubuntu Linux 22.04.

As soon as the system is powered on, it continuously begins to recognize ArUco markers. The cameras detect 5x5-sized ArUco markers placed in front of each shelf. If the ``\textit{Section Description}" button is pressed, the name of the section associated with the marker is read aloud to the user. Since supermarkets have many shelves, to avoid overlapping commands, the system applies two criteria for reading a marker: the first is the marker’s size, which varies depending on the distance from the cart when the image is captured, larger markers (closer) have higher relevance. The second factor is the time since the marker was last read; a marker is only re-read after a certain interval or after a different marker has been recognized. After being recognized, the last marker ID from the left and right cameras is stored in two separate variables to help determine the user’s location and orientation. However, if no new marker is recognized after a few seconds, the variable is cleared to avoid false localization.

In contrast to continuous marker detection, photographic image capture for environmental analysis is performed on-demand. This occurs only when the user presses either the \textit{Environment Description Button} or the \textit{Voice Command Button}. In such cases, the Image Capture Module simultaneously captures three images — one from each camera — representing the left, front, and right views.

These captured images are then transmitted to the Processing Module, where they are analyzed to generate environment descriptions or to assist in answering user queries via the integrated AI system.

\subsection{Audio Capture Module}
The Audio Capture Module is responsible for acquiring and processing user voice commands, which form a core part of the system’s interactive capabilities. This module enables users to query product information, request environment descriptions, or initiate navigation through simple verbal instructions.

When the user presses the \textit{Voice Command Button}, the Audio Capture Module is activated. At this point, the system begins recording audio through an onboard microphone. To efficiently manage the recording process and avoid unnecessary data capture, the module employs a \textit{Voice Activity Detection (VAD)} mechanism. Specifically, the WebRTC Voice Activity Detector (webrtcvad) is used to continuously analyze short frames of audio in real time. The VAD determines when speech is present and automatically stops the recording once speech ceases, signaling that the user’s utterance has concluded.

The recorded audio is then transcribed into text using the {Whisper-1} automatic speech recognition (ASR) model provided by OpenAI. This model is equivalent to the Large-v2 version of Whisper and contains approximately 1.55 billion parameters, offering robust transcription capabilities for diverse speech patterns and environments.

Once transcription is complete, the resulting text is forwarded to the \textit{Processing Module} for further interpretation. Depending on the nature of the user’s request, the Processing Module may perform semantic queries in the local database, initiate navigation instructions, or, if needed, activate the Image Capture Module to support context-aware responses.

By integrating VAD and ASR technologies, the Audio Capture Module provides an intuitive and streamlined mechanism for users to interact naturally with the system without requiring visual input or complex operations.


\subsection{Question and Data Processing Module}
The Question and Data Processing Module serves as the central decision-making component of the system, interpreting user commands and coordinating subsequent actions. Whether initiated through voice commands or environment description requests, this module determines the appropriate response path based on the nature of the user’s query.


Upon activation, the module processes inputs received from either the Audio Capture Module (in the form of transcribed text) or the Image Capture Module (as visual data). For voice-triggered queries, the transcribed command is first analyzed to classify the user’s intent into one of the following categories:
\begin{itemize}
\item \textbf{Product Information Retrieval:} Queries related to the location or price of products.
\item \textbf{Navigation Requests:} Commands that require guiding the user to a particular product or store section.
\item \textbf{General Environment or Contextual Queries:} Requests for broader information regarding the surroundings or requiring scene analysis.
\end{itemize}
If the command relates to product information, the system performs a \textit{semantic search} within the local product database. To achieve this, the transcribed user query is converted into a high-dimensional vector representation using the \textit{text-embedding-ada-002} model from OpenAI. This vector is then compared with precomputed vectors for each product entry in the database using {cosine similarity}, as defined in Equation~\ref{eq:cosine}:
\begin{eqnarray}
\text{similarity}(A, B) = \frac{A \cdot B}{|A| |B|}
\label{eq:cosine}
\end{eqnarray}

\noindent where A and B represent the embedding vectors of the user query and the database entry, respectively, and $\|\cdot\|$ denotes the Euclidean norm.
If the computed similarity exceeds a threshold of 90\%, the system considers this a precise match and directly announces the product’s location (including shelf ID, vertical and horizontal shelf position) and price to the user. If the similarity falls below this threshold, the system retrieves and presents the top three most similar candidates, offering the user a choice to select the correct product.

When a user request pertains to navigation, the Processing Module engages the Navigation Module. It provides the current shopping cart location (based on ArUco marker detection) and the desired destination (obtained from the product database). The Navigation Module then generates a step-by-step route to guide the user to the target location.

If the user’s request falls outside the scope of product queries or navigation—for example, general questions about the surroundings or descriptions of the environment—the Processing Module forwards both the transcribed user input and any captured images to the integrated AI model for open-ended reasoning and natural language generation.

In this system, \textit{ChatGPT 4o}, a recent and advanced version of the GPT (Generative Pre-trained Transformer) family, is utilized for this purpose. ChatGPT 4o is optimized for real-time dialogue, multimodal understanding, and context-aware reasoning tasks. Unlike base GPT models, ChatGPT 4o maintains conversational coherence, processes multimodal inputs (including images), and supports function calling mechanisms, making it especially suitable for dynamic and natural communication in assistive scenarios such as this\cite{openaiOpenAI}.

Once ChatGPT 4o generates a contextually appropriate response, the system converts this text output into audio through the Speaking Module and delivers it to the user. Following the completion of any processing path, the system returns to the \textit{Listening Module} to await the next user command.

\subsection{Navigation Module}

The navigation module provides optimized guidance to the user based on the location identified through continuous recognition of ArUco markers placed throughout the environment. Detecting these markers with the system’s cameras makes it possible to determine both the user’s current position on the supermarket map and their initial movement orientation, depending on whether the left or right camera detected the marker.

For route planning, the A* algorithm is used, known for its efficiency in finding optimal paths in graphs. It calculates the total cost f(n) of each node based on the sum of the actual cost g(n) from the starting point and a heuristic h(n) that estimates the cost to the destination, as shown in Eq. (\ref{eq:a_star}):
\begin{eqnarray} 
\begin{array}{rcl} f(n) &=& g(n) + h(n) 
\end{array} 
\label{eq:a_star} 
\end{eqnarray}
In this system, the algorithm's inputs are the initial position (obtained from the recognized ArUco marker), the target position (e.g., desired section or product), and the initial orientation (based on which camera identified the marker).

\begin{figure}[!t]
\begin{center}
\includegraphics[width=8.5cm]{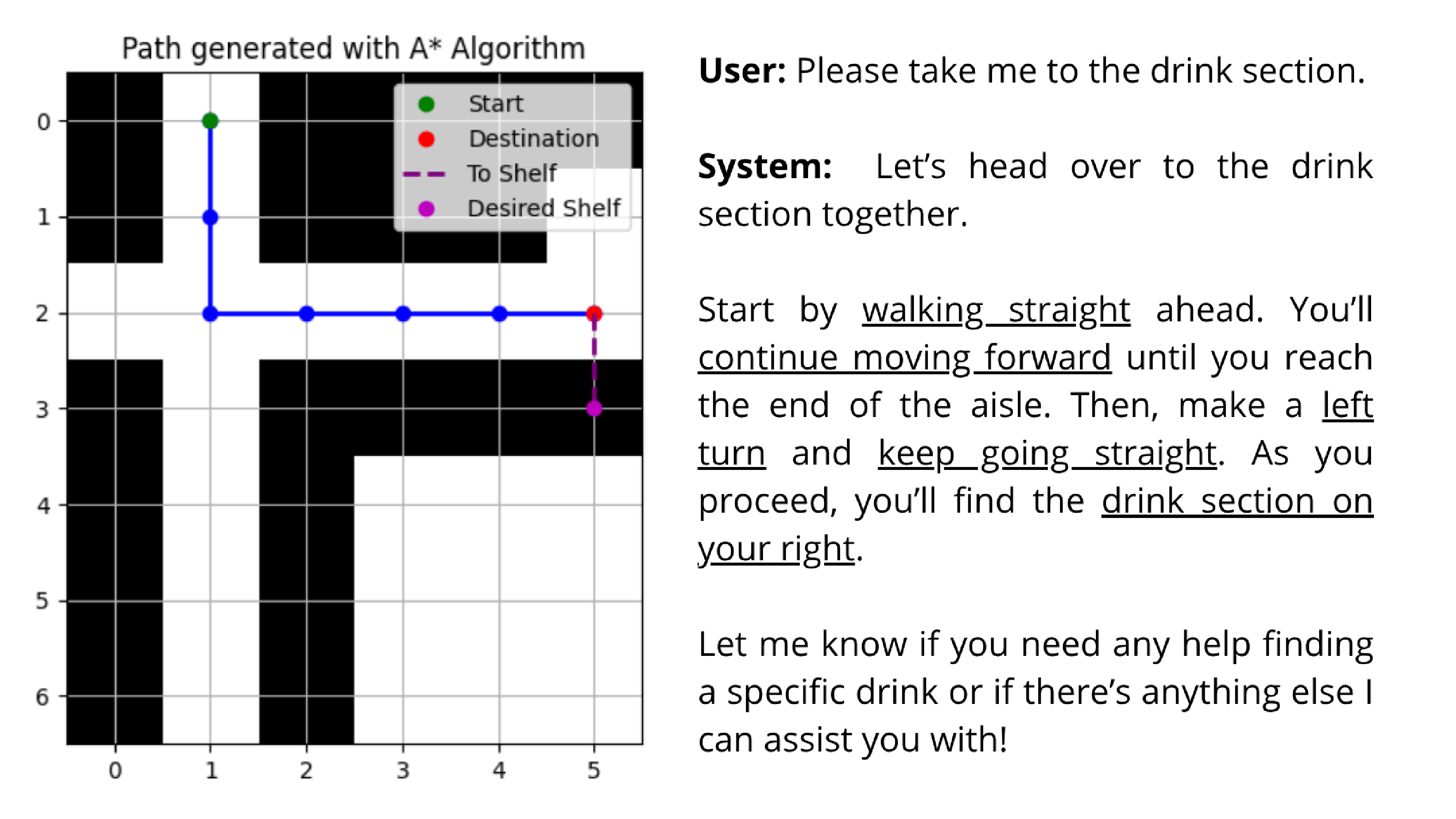}
\caption{\label{algorithm} {Example of a route generated by the A* algorithm and the corresponding dialogue between the user and the system. The visualized path represents the optimal trajectory from the user's current location to the desired shelf, while the dialogue illustrates how the system translates this route into natural language instructions for a visually impaired user.}}
\end{center}
\end{figure}

The module’s output is an optimized path, represented as a sequence of movements on the supermarket map. To make navigation accessible and understandable for visually impaired individuals, this sequence of movements is converted into natural language instructions. This conversion is performed using a GPT-based model, which interprets the sequence of directions and generates clear, humanized audio commands such as “turn right,” “go straight for two aisles,” or “turn left at the next section.”
{An example of an interaction with the system can be seen in Figure~\ref{algorithm}.}

\subsection{Speaking Module}

Finally, the speaking module is responsible for converting the responses generated by the AI into natural-sounding audio output, enabling auditory interaction with the user. This module is essential for providing feedback to visually impaired users in a seamless and intelligible manner.

For this task, the system employs the \textit{OpenAI TTS} (Text-to-Speech) tool with the Onyx voice. After processing the request, the textual response is broken into smaller segments, especially in cases of longer messages, to reduce the perceived latency for the user. Each segment is individually converted into audio and stored in a playback queue, ensuring that the information is delivered in a smooth and efficient manner.

In addition to text-to-speech conversion, the system incorporates non-verbal auditory feedback at strategic moments during the interaction. A brief sound is played to indicate the start of voice recording in the audio capture module, informing the user that the system is actively listening. Another sound is emitted while the AI processes the response, signaling that the system is working on an answer. These audio cues enhance the interactivity and transparency of the system’s behavior, especially for users with visual impairments and who rely entirely on auditory perception.

All synthesized speech and sound cues are delivered through a headphone output, allowing the user to receive private and uninterrupted feedback even in noisy store environments. Once the message or instruction is fully delivered, the system transitions back to the \textit{Listening Module}, ready to accept the next input.

\section{Experimental Results and Discussion}

\begin{figure}[!b]
\begin{center}
\includegraphics[width=8cm]{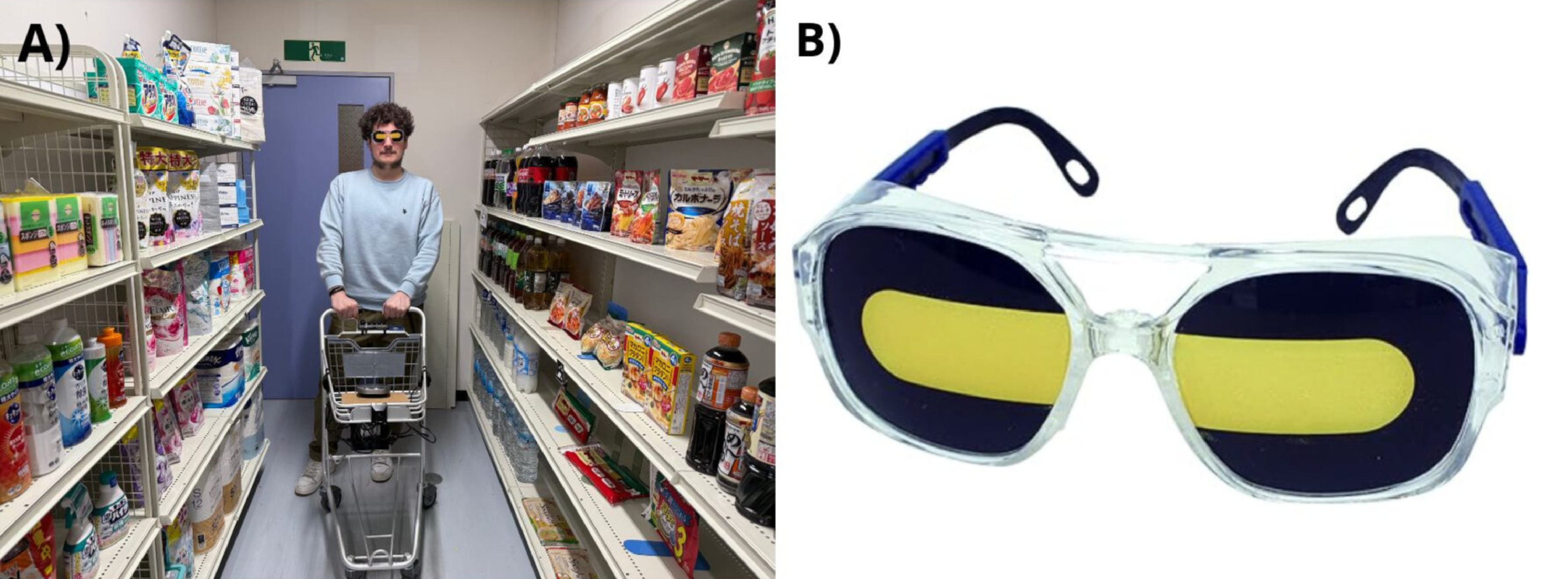}
\caption{\label{test}(A) Experimental setup with the AI shopping assistant cart during tests. (B) Cataract simulation goggles used to reproduce the visual impairment condition during the evaluation.}
\end{center}
\end{figure}

Two experiments were conducted to evaluate the usability and effectiveness of the proposed system under normal and impaired vision conditions. The experiments were designed to assess the system’s ability to support in-store navigation and product detection tasks in a realistic shopping scenario. Each experiment consisted of two phases: one performed under normal vision, and another using a cataract-simulation device (see Fig. \ref{test}) to mimic moderate to severe visual impairment.

Both experiments were conducted with four participants, two of whom had received prior training with the system. To evaluate cognitive and physical workload, participants completed the {NASA Task Load Index (NASA-TLX)} questionnaire after each phase. The NASA-TLX assesses subjective workload across six dimensions: \textit{Mental Demand}, \textit{Physical Demand}, \textit{Temporal Demand}, \textit{Performance}, \textit{Effort}, and \textit{Frustration}. Each dimension is rated from 0 (low) to 100 (high), and the weighted sum provides an overall workload score. Higher scores indicate increased perceived workload or emotional strain\cite{hart2006nasa}.

\subsection{Navigation Task Results}

In this experiment, the participants were instructed to locate three store sections within a simulated supermarket. The requested sections were not visually associated with nearby products, ensuring that participants relied on system guidance. Different section targets were assigned across conditions to prevent memory bias. After each phase, participants completed the weighted NASA-TLX questionnaire.
Results for the navigation task are summarized in Table \ref{tab:navigation_results}.

\begin{table}[htb]
\small
\caption{\label{tab:navigation_results}Navigation Task: Summary of results.}
\begin{center}
\begin{tabular}{|c|c|c|c|c|}\hline
\textbf{Condition} & \textbf{Normal Vision} & \textbf{Impaired Vision} \\\hline
Success Rate & 91.6\% & 100\% \\\hline
Avg. Time (s) & 59.6 & 57.4 \\\hline
Avg. Queries  & 4.3  & 3.75\\\hline
NASA-TLX score  & 35.9 & 40.4 \\\hline
\end{tabular}
\end{center}
\end{table}

Although participants with normal vision could theoretically rely on visual cues, all information was intentionally hidden or unlabeled to force dependence on the AI system. Interestingly, one participant failed to reach a target section due to incorrect system feedback. This incident, combined with initial unfamiliarity, likely explains why the average success rate and NASA-TLX score appear worse under normal vision than impaired vision. This highlights how prior expectations and the mismatch between visual input and auditory feedback can increase cognitive load.

Among trained participants, the average number of queries was lower (3.0) compared to untrained participants (5.0),  and the time to reach each section was shorter (73 s vs. 137 s), demonstrating a learning effect. However, the Frustration score was paradoxically higher among trained users (277.5 vs. 45). This may reflect raised expectations and increased sensitivity to system limitations.

In this experiment, Section 1 took the longest time to be located (105 seconds on average) and required the highest number of queries to the system (2.25). NASA-TLX scores revealed that Frustration and Performance were the most influential dimensions, with average weighted scores of 161.3 and 116.3, respectively.

Under simulated visual impairment, all participants completed the task successfully. Fewer queries were needed, and the average time was slightly lower. These results likely reflect both the absence of conflicting visual cues and increased familiarity with the system by the second phase.

For impaired vision, the most influential NASA-TLX dimensions were Performance (150.0) and Effort (135.0). These results suggest that visual impairment led participants to perceive the task as more cognitively demanding, requiring greater effort to achieve satisfactory performance.

Voice system latency (measured from the voice button press to audio output) averaged 19.3 seconds under normal vision and 20.9 seconds under impaired vision. This includes time for audio recording, transcription, processing, and text-to-speech synthesis. A total of six system errors or unexpected behaviors were reported, which likely contributed to increased task duration or failure in some cases.  

\subsection{Product Detection Task Results}
In the second experiment, participants were asked to find two products on store shelves using voice queries. The product names varied across phases and were generic (e.g., "instant noodles"), allowing the participant to choose among system-suggested options. The task was performed with and without the vision impairment simulation, and participants completed a NASA-TLX questionnaire after each phase.
Results for this experiment are summarized in Table \ref{tab:detection_results}.

\begin{table}[htb]
\small
\caption{\label{tab:detection_results}Product Detection Task: Summary of results.}
\begin{center}
\begin{tabular}{|c|c|c|c|c|}\hline
\textbf{Condition} & \textbf{Normal Vision}  & \textbf{Impaired Vision} \\\hline
Success Rate& 87.5\% & 100\%  \\\hline
Avg. Time (s) & 143.8 & 136.1  \\\hline
Avg. Queries & 9.25 & 7.25  \\\hline
NASA-TLX score & 50.7 & 44.8  \\\hline

\end{tabular}
\end{center}
\end{table}

In the normal vision condition, one participant failed to locate a product. On average, more queries were needed (9.25), especially for the first item (6.25 queries), which took significantly longer to find (218 seconds vs. 69.5 seconds for the second). This suggests a learning curve during interaction.

Despite the availability of visual input, participants found the system difficult to use at first. The discrepancy between what they saw and what the system communicated may have caused frustration, reflected in high NASA-TLX Frustration (243.8) and Mental Demand (183.8) scores.

With simulated vision impairment, performance improved—likely due to increased system familiarity and focused reliance on auditory input. NASA-TLX scores showed highest weights on {Frustration} (172.5) and {Effort} (157.5). Voice latency averaged 18.0 seconds (normal vision) and 17.1 seconds (impaired).

Voice latency was slightly lower than in the navigation task, averaging 18.0 seconds (normal) and 17.1 seconds (impaired). Seven system errors occurred, often related to Whisper transcription issues.

\subsection{NASA-TLX Analysis}

\begin{figure}[h]
\begin{center}
\includegraphics[width=8cm]{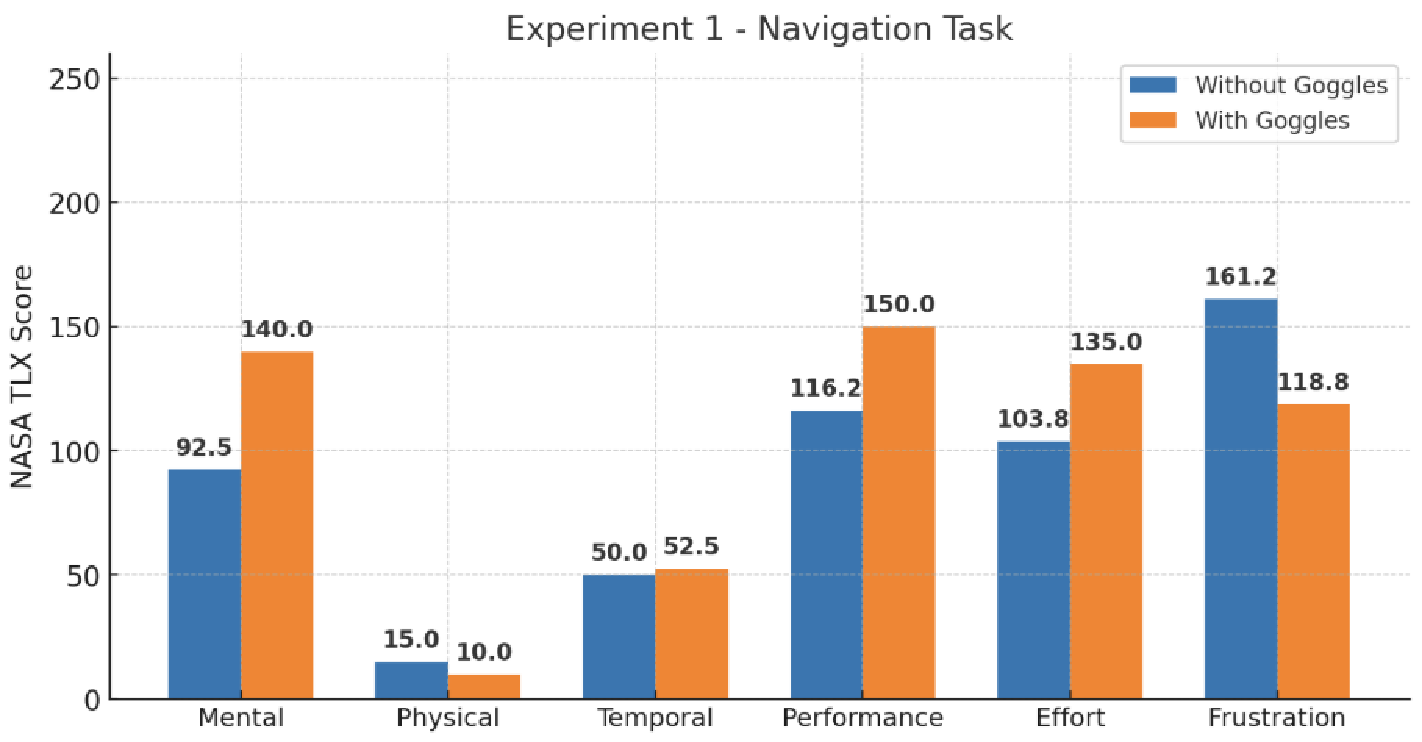}
\caption{\label{figure3} Nasa TLX scores per dimension - Navigation Experiment.}
\end{center}
\end{figure}

\begin{figure}[h]
\begin{center}
\includegraphics[width=8cm]{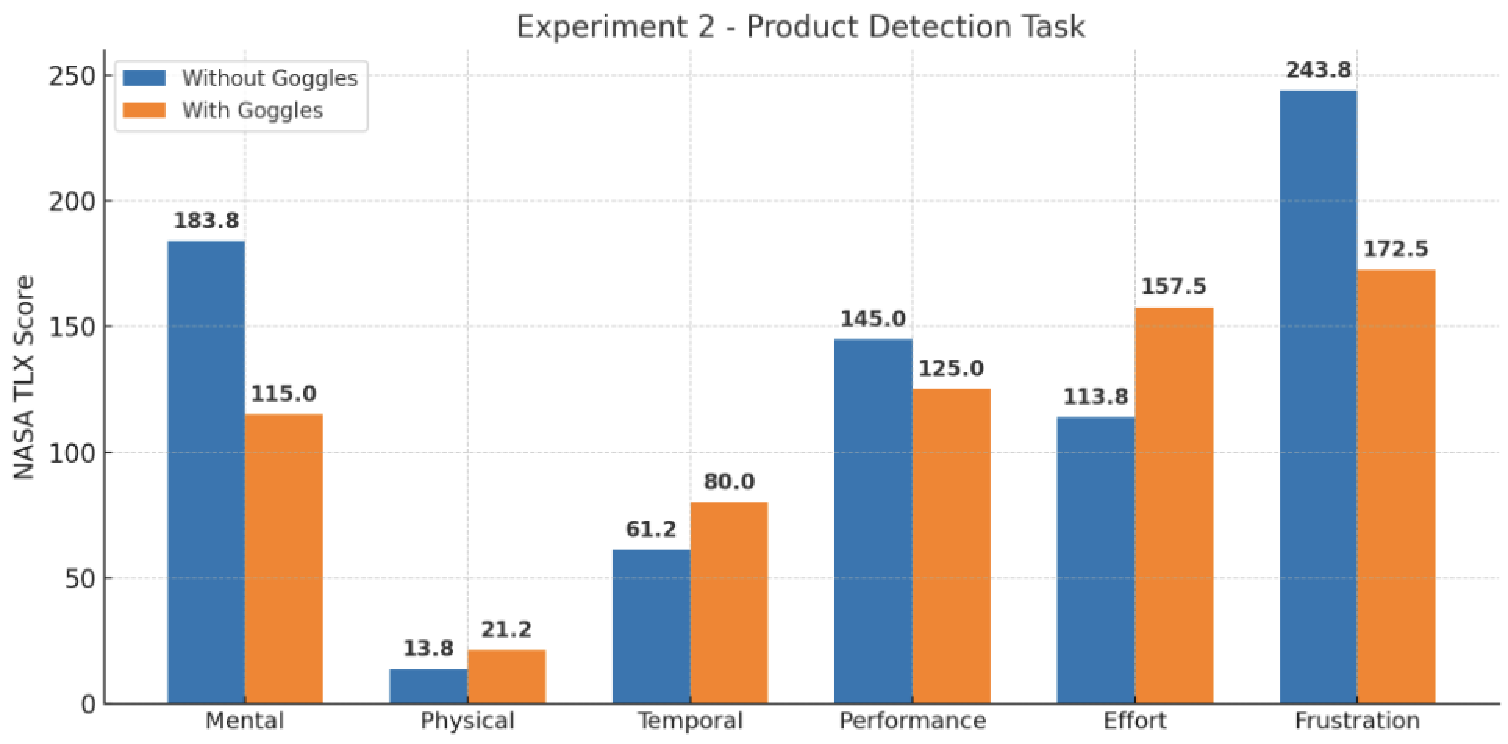}
\caption{\label{figure4} Nasa TLX scores per dimension - Product Recognition Experiment.}
\end{center}
\end{figure}

A summary of weighted NASA-TLX scores per dimension is presented in Figure \ref{figure3} and Figure \ref{figure4} for each experiment. 

{Frustration} and {Perceived Performance} were the most influential dimensions, especially during early interactions or when user expectations were not met. Over time, participant familiarity with the system led to reduced workload scores and increased efficiency.


\section{Conclusion and Future work}

This work presented an AI-based shopping assistant designed to support blind and visually impaired individuals by promoting greater autonomy during grocery shopping. Implemented as a smart shopping cart, the system integrates vision, speech, and language technologies to guide users, identify products, and interact conversationally.

Experimental evaluations conducted under both normal and simulated vision impairment conditions confirmed the system’s ability to assist users in navigating the shopping environment and locating products, with improved performance through training. However, delays in response generation were observed due to processing time. To address this, faster local modules like faster-whisper and piper will be adopted. Improvements in navigation instruction clarity and user localization are also planned.

Future efforts include testing with visually impaired participants in real supermarkets to validate usability and expand functionality in real-world conditions.

\section{Acknowledgments}

This work was partially supported by JST [COI-NEXT][Grant Number JPMJPF2201] and JSPS Kakenhi [Grant Number JP24K07399].

\bibliographystyle{ieeetr}
\tiny
\bibliography{SICE-FES_Rev_v2.bib}

\end{document}